# Suppression of single-cesium-atom heating in a microscopic optical dipole trap for demonstration of an 852-nm triggered single-photon source

Bei Liu[1, 2], Gang Jin[1, 2], Jun He[1, 2, 3], and Junmin Wang[1, 2, 3, *]

1. Institute of Opto-Electronics, Shanxi University, Tai Yuan 030006, Shan Xi, P. R. China

2. State Key Laboratory of Quantum Optics and Quantum Optics Devices, Shanxi University, Tai Yuan 030006, Shan Xi, P. R. China

3. Collaborative Innovation Center of Extreme Optics, Shanxi University, Tai Yuan 030006, Shan Xi, P. R. China

**Abstract:** We investigate single-cesium-atom heating owing to the momentum accumulation process induced by the resonant pulsed excitation in a microscopic optical dipole trap formed by a strongly focused 1064-nm laser beam. The heating depends on the trap frequency, which restricts the maximum repetition rate of pulsed excitation. We experimentally verify the heating of a single atom and then demonstrate how to suppress it with an optimized pulsed excitation and cooling method. The typical trap lifetime of single cesium atom is extended from 108 ± 6 μs to 2536 ± 31 ms, and the corresponding number of excitations increases from ~ 108 to ~ 360000. In applying this faster cooling method, we use the trapped single cesium atom as a triggered single-photon source at an excitation repetition rate of 10 MHz. The second-order intensity correlations of the emitted single photons are characterized by implementing Hanbury Brown and Twiss setup, and clear anti-bunching effect has been observed.

## I. INTRODUCTION

Atom-photon interactions can be used to demonstrate photonic and atomic qubits for quantum information processing. Recently, the study of atom–photon interactions based on trapped neutral atoms has made significant progress [1-5]. A two-level neutral atom system is generally used to generate single photons or implement a qubit with long coherence times. Single-qubit or multiqubit operations have been realized by using microwaves to drive the hyperfine transition directly or by using a two-photon Raman transition [6-9]. Pulsed excitation of a two-level atom is also a promising tool for effective generation of narrowband single photons [10,11]. Interference between the two photons has been demonstrated by using the Hong-Ou-Mandel (HOM) interference experiment [12,13]. Additionally, preparation of atoms in Rydberg states can control the entanglement between the atoms by use of the Rydberg blockade effect [14,15]. As in most of the above applications, the qubit operation and single-photon generation require control of the interaction between the excitation laser and the atoms. However, the interaction between the frequency-detuned Raman pulsed laser and trapped single atom will heat up the atoms, resulting in decoherence. The interaction between a

---

near-resonance pulsed laser and a single atom causes accumulation of atomic momentum, leading to loss of atoms from a dipole trap [10].

The knowledge and suppression of the heating of a single atom, after pulsed excitation, is important for many purposes. For instance, in a qubit operation experiment [6,7], the qubit rotation is performed by a frequency detuned Raman pulsed laser. The heating induced by the pulsed Raman laser accelerates the atomic motion, leading to a differential light shift of the clock transition, which results in inhomogeneous dephasing. For this application, suppression heating of the trapped atom would increase the dephasing time, thus avoiding the extra "spin-echo" sequence. Elsewhere, a single atom that is excited with high-repetition-rate and short resonant light pulses has been used to demonstrate a triggered single-photon source [10,11]. The pulsed excitation causes accumulation of momentum, leading to loss of the atom in less than a millisecond. Furthermore, the repetition rate of single-photon generation is limited by the atom heating depending on the trap frequency. Thus, for a triggered single-photon source, suppression heating of the trapped atom would increase the pulsed excitation time and create a maximum number of photons.

In this paper, we analyze in more detail the heating mechanism and the cooling methods to further extend the pulsed excitation time of the atom. The paper is organized as follows. In Sec. II we theoretically analyze the heating mechanism which depends on the parameters of the trap. The experimental setup is described in Sec. III. We measure the heating induced by the pulsed laser in Sec. IV, where it is shown that the heating depends on the repetition rate of pulsed laser and the parameters of the optical dipole trap. In this section, the heating is suppressed with the assistance of a gated pulse excitation and cooling method [16, 17], an optimized timing sequence, and appropriate parameters for the cooling laser. This technique enables pulsed excitation with a high repetition rate up to 10 MHz. Finally, we measure the photon correlation functions of single-atom emission by implementing the Hanbury Brown-Twiss (HBT) setup [18], and clear antibunching effect is observed.

## II. THEORETICAL ANALYSIS

The heating mechanisms for atoms in an optical dipole trap include the momentum accumulation from pulsed excitation, collision with the background gas, and parametric heating from the trap laser intensity or frequency fluctuation [16, 19]. In our experiment, the typical trap lifetime (without pulsed excitation) has been improved from ~ 6.9 s to ~ 130 s by decreasing the background pressure from ~ 1 × $10^{-10}$ Torr to ~ 2 × $10^{-11}$ Torr and applying 10-ms cooling phase [19]. Thus, for a well-stabilized optical dipole trap in an ultrahigh-vacuum chamber [20], the heating induced by collisions and the trap laser is negligible relative to that induced by the pulsed laser.

We first discuss the heating induced by the pulsed laser. In the following calculations, a single atom in an optical dipole trap will be treated as a classical three-dimensional harmonic oscillator. The single atom remains initially at the bottom of the trap when it is trapped. This assumption is justified because the expected temperatures of the atom is 17 μK [19], which is ten times lower than the trap depth $U_0$ ~ 2 mK. When a single atom is excited by a pulsed laser, the induced heating can be studied for two cases [21]. In the case where $\Omega_{Pulse} < \Omega_{Trap}$ ($\Omega_{Pulse}$ is the repetition rate of the pulsed laser and $\Omega_{trap}$ is the trap frequency), the atom oscillates many times while absorbing and emitting photons; hence, the momentum transfer averages to zero. Each absorption and emission process increases the atomic energy by $E_r = \hbar^2 k^2 / 2m$. In the case where $\Omega_{Pulse} > \Omega_{Trap}$, the high repetition rate of excitation process causes accumulation of the photon momentum ($p = \hbar k$), leading to an increase of the atomic total energy. To lose an atom from a dipole trap with depth $U_0 \sim 2$ mK, the recoil heating and the accumulation all photon momentum processes need the number of pulsed excitation $n_1 = 10152$ and $n_2 = 142$, respectively, where the total energy $E_0 = 2n_1 E_r$ and $E_0 = (n_2 p)^2 / 2m = n_2^2 E_r$, the following relation holds $n_2 = \sqrt{2n_1}$.

During pulsed excitation, an atom has the possibility to escape from the trap if its total energy $E_0$ is larger than the trap depth $U_0$. It is assumed that the atoms have a thermal distribution $D_s(E)f_{th}(E)$ inside the trap which is defined by its temperature [22]. The temperature increases linearly with each pulsed excitation process, given by $T(t) = T_0 + \alpha t$, where $T_0$ is the initial temperature and $α$ is the heating rate, $t$ is the trap lifetime. The behavior of the measured recapture probability gives us the trap lifetime of single atom. It can be calculated by integrating the energy distribution $P(t) = \int_{U_0}^0 D_S(E) f_{th}(E) dE = 1 - \left[1 + \eta + \frac{1}{2}\eta^2\right] e^{-\eta}$, $\eta = U_0 / (k_B T(t))$.

We then analyze the dependence of the heating rate on the parameters of the trap. In our system, the repetition rate of the pulsed laser is 10 MHz, the pulsed excitation process causes accumulation of photon momentum, leading to the atom being expelled from the trap. As the trap has two different trap frequencies along the axial and radial directions, these corresponds to two different maximum repetition rates. For a 2-mK trap depth, the maximum number of pulsed excitation is $n_2 = 142$, and the calculated trap frequency in the axial and radial directions are $\Omega_{axial} = 2\pi \times 4.9 kHz$ and $\Omega_{radial} = 2\pi \times 47.8 kHz$. The estimated maximum pulse repetition rate is 0.7 MHz in the axial direction and 6.8 MHz in the radial direction. In this case, the repetition rate of single-photon generation is restricted by the atom heating and depends on the trap frequency. Finally, a gated excitation and cooling method can be used to suppress the heating and break the repetition rate restriction by faster laser cooling. The gated pulse excitation-cooling cycle means that the pulse excitation beam and the cooling beam are switched on and off alternately. Thus, after each pulsed excitation, the atom is rapidly cooled to the bottom of the trap by the polarization gradient cooling (PGC) with an "σ $^+$- σ $^-$" configuration. The PGC phase is provided by the magneto- optical trap (MOT) cooling and repumping laser beams [23].

### III. EXPERIMENTAL SETUP

A schematic of experimental setup is shown in Fig. 1. The related setup has been described in detail elsewhere [19, 20]. An objective lens with a high numerical aperture (0.29) produces a diffraction-limited beam waist of 2.3 μm. The trap depth is about 2 mK for a 1064 nm laser power of ~ 63 mW. The 852-nm fluorescence photons from the trapped atom are collected by the same objective lens. The light-induced fluorescence is separated from the trapping laser by a dichroic mirror. The bottom right panel depicts the fluorescence photon counting signal from a single atom, which is continuously being loaded for over 500 s. The trapped single atom is excited by the MOT laser beams with an average trap lifetime of about 12 s. To prove the single-photon characteristics of our source, we use a standard HBT setup (shown at the bottom left). This setup is composed of two single-photon-counting modules (SPCM1 and SPCM2) behind a 50 : 50 beam splitter (BS). The electrical pulses from the SPCMs are sent to a P7888 card (two-input multiple-event time digitizers, FAST Com Tech.) for time-resolved analysis. Using this HBT setup, we measured the second-order correlation function of the fluorescence photons emitted by the trapped atom under continuous-wave laser excitation, and found $g^{(2)}(\tau=0) = 0.08$ [24].

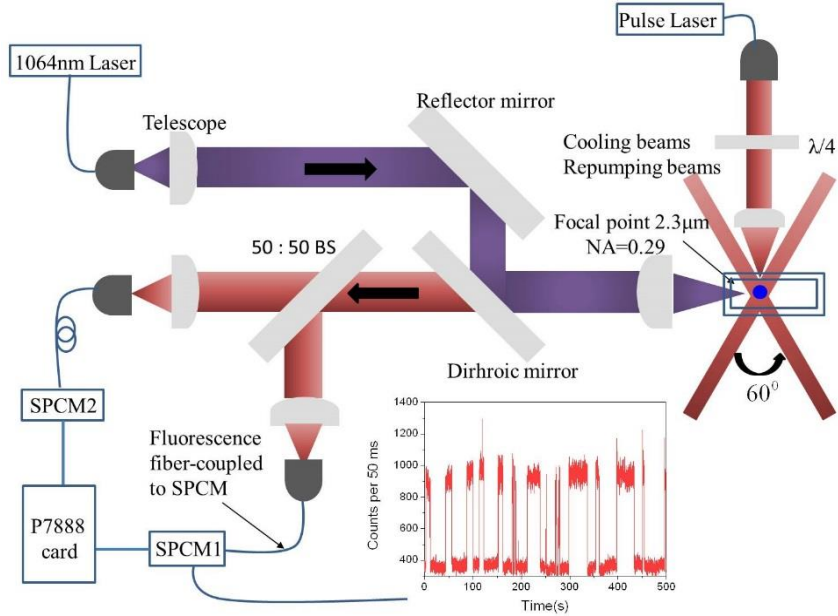

**FIG. 1** Experimental setup. A single atom is trapped in an optical dipole trap (bottom right). The light-induced fluorescence is separated from the trapping laser by a dichroic mirror. The fluorescence photons from the single atom are sent through a multi-mode fiber to the HBT setup (bottom left), which consists of a 50:50 beam splitter and two SPCMs.

The trapped single atom is excited by the pulsed laser in the radial direction of the dipole trap. The pulsed laser is generated by directing a continuous-wave laser through an electro-optical intensity modulator (EOIM, EO Space). By optimizing the polarization of the incident laser beam and by stabilizing the temperature of the EOIM, the extinction ratio is improved to 12600:1[25]. The EOIM's extinction ratio is high enough to avoid pulsed excitation during the off phase[11]. To match the ac-Stark shift of the Cs $6S_{1/2}$ |Fg = 4, $m_F$ = +4> - $6P_{3/2}$ |Fe = 5, $m_F$ = +5> cycling transition induced by the optical trap laser, the frequency of the pulsed laser is shifted with an acousto-optical modulator (AOM). The excitation beam waist is ~ 12 μm at the location of the atom. The pulsed laser beam is $\sigma^+$-polarized relative to the quantization axis by using of a Glan-Taylor prism and a quarter-wave plate. The quantization axis is defined by a 0.2-mT (2-Gauss) magnetic field along the radial direction.

## IV. EXPERIMENTAL RESULTS AND DISCUSSION

Atom heating owing to the momentum accumulation process induced by the pulsed excitation depends on the repetition rate of pulsed laser and the trap frequency. In this section, the heating is determined by measuring the lifetime of the trapped single atom after pulsed excitation. To suppress the heating, a gated excitation and cooling technique is then applied. With the optimized parameters of the cooling laser, the atoms can be repeatedly excited for long times at a high repetition rate. Finally, using this optimized excitation and cooling sequence, we are able to use the single trapped atom to generate triggered single photons.

Fig. 2 shows a schematic of the experimental sequence, which goes as follows: (i) The single atom is trapped and laser cooled in the MOT with a large-magnetic-field gradient. (ii) The MOT is turned off and the optical dipole trap is turned on; the single atom is transferred into the optical dipole trap with 10-ms PGC phase. (iii) The single atom is illuminated by resonant unidirectional σ+-polarized pulsed laser. In the pulsed excitation process, the quantification magnetic field is applied continuously and the repumping beams are always on to avoid the atom depumping into the $F_g = 3$ ground state. In order to improve the lifetime under pulsed excitation, the atom is alternately excited and cooled. The detection gate is used to open the detection window during pulsed excitation. Finally, the MOT is switched back on to determine if the atom has escaped or not.

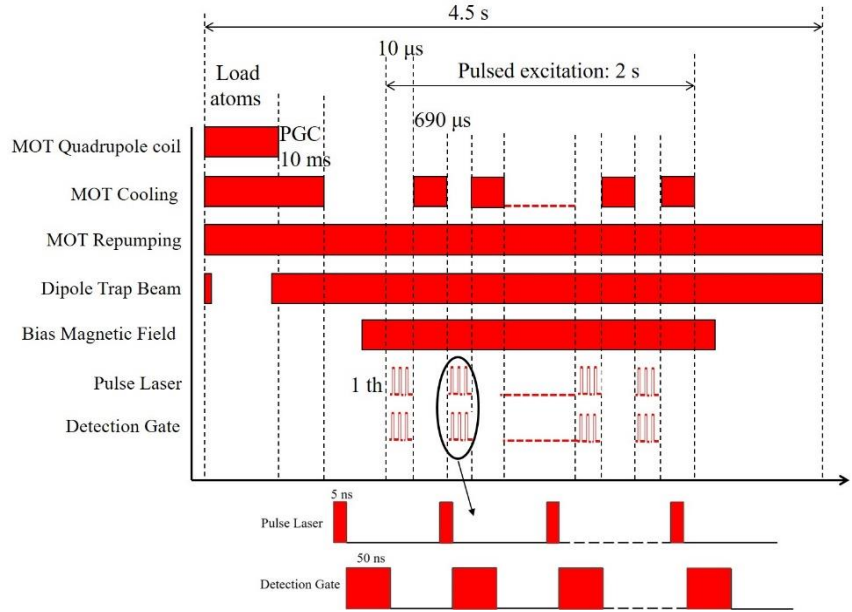

**FIG. 2** Experimental timing sequence. A single atom is confined in an optical dipole trap and then excited by a resonant pulsed laser. A gated excitation/cooling technique is employed to suppress the heating. The fluorescence signal is gated so that only photons emitted during the periods of pulsed excitation is counted.

To measure the single-atom lifetime, we use the experimental sequence illustrated in Fig. 2. The single atom is trapped in an optical dipole with $U_0 \sim 2mK$. Its temperature is measured to be $T_0 = 17$ μK initially [19]. Then the atom is excited for a time $t_{ext}$ and we check if the atom is still existed afterwards. Over 100 repetitions we measure the recapture probability ($P_R$) in dependence on the excitation time $t_{ext}$. As shown in Fig. 3 (a), the single atom is excited at different repetition rates in the radial direction. The diamonds and circles data points are for repetition rates of 30 kHz and 1 MHz, respectively. Each data point is the accumulation of 100 sequences, with error bars due to the binomial statistics. We fit the function $P_R = P(t)\exp(-t/\tau)$ to the data and obtain the trap lifetime $\tau = 163 \pm 18$ ms (for 30 kHz) and $\tau = 108 \pm 6$ μs (for 1 MHz), respectively. The number of the allowed excitation is $n_{ext} \sim 4890$ (30 kHz) and $n_{ext} \sim 108$ (1 MHz). The theoretical values of $n_{ext} \sim 10152$ (30 kHz) and $n_{ext} \sim 142$ (1 MHz) are higher than the experimental results; this can be explained by the fact that the heating can also be induced by the collision with the background gas and parametric heating from the trap laser intensity or frequency fluctuation.

Fig. 3 (b) shows the excitation of in different excitation directions with 10 MHz repetition rate.

The diamonds data points show the radial excitation and the circles points show the axial excitation. The atom is excited for 10 μs, separated by waiting periods of 690 μs. The trap lifetimes are τ = 52.3 ± 4.1 ms (the radial direction) and τ = 5.2 ± 0.4 ms (the axial direction). This result shows that the atom has stronger confinement in the radial direction compared with the axial direction. In the radial direction, the atom can be excited with a higher repetition rate.

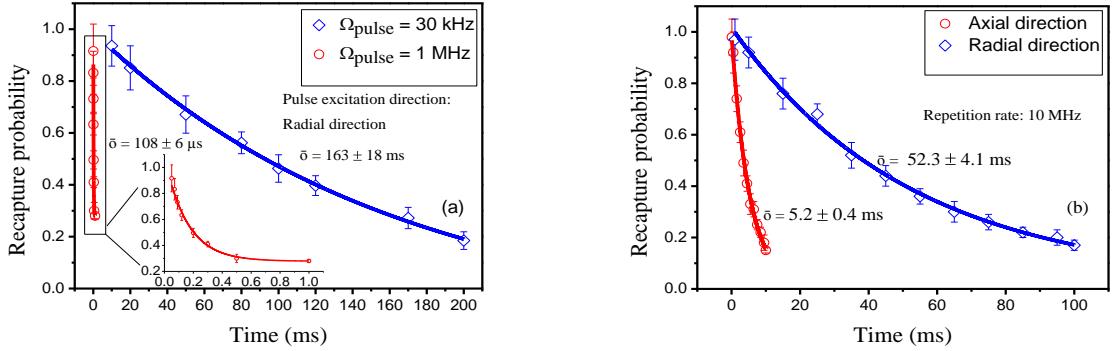

**FIG. 3** (a) shows that the atom is excited at different repetition rate with the same excitation direction. Inset shows zooming of the data for 30 kHz. (b) shows that the atom is excited at different excitation direction with a repetition rate of 10MHz. All the data point is the accumulation of 100 sequences, with error bars due to the binomial statistics. The solid line is theoretical fitting to the data by P(t). The result shows that the atom can be excited with a high repetition rate in the radial direction. Experimental parameters: pulse duration: 5 ns, pulse laser beam's power: 1.25 mW, pulse laser beam's waist: ~ 12 μm.

Using the gated excitation and cooling sequence, we are able to significantly reduce the heating and extend the lifetime. The trap lifetime is optimized by controlling the duration and intensity of the cooling laser. Fig. 4(a) shows the dependence of the trap lifetime on the intensity of the cooling laser, when the atom is excited for 10 μs and cooled for 4900 μs and the detuning of the cooling laser is -8 Γ (Γ = 2π × 5.2 MHz is the natural linewidth of the Fe = 5-Fg = 4 hyperfine transition). The results indicate that the trap lifetime increases as the cooling laser intensity increases from 0 to 12 $I_{sat}$. Fig. 4(b) shows the dependence of the trap lifetime on the PGC interaction time. The single atom is excited for 10 μs and the PGC interaction time varied over 0-5000 μs. The cooling laser intensity is 10 $I_{sat}$ and the detuning is -8 Γ. The trap lifetime increases rapidly from ~ 56 ms to ~ 2536 ms when the cooling time increases from 190 μs to 690 μs, and then it reaches a steady value after 690 μs.

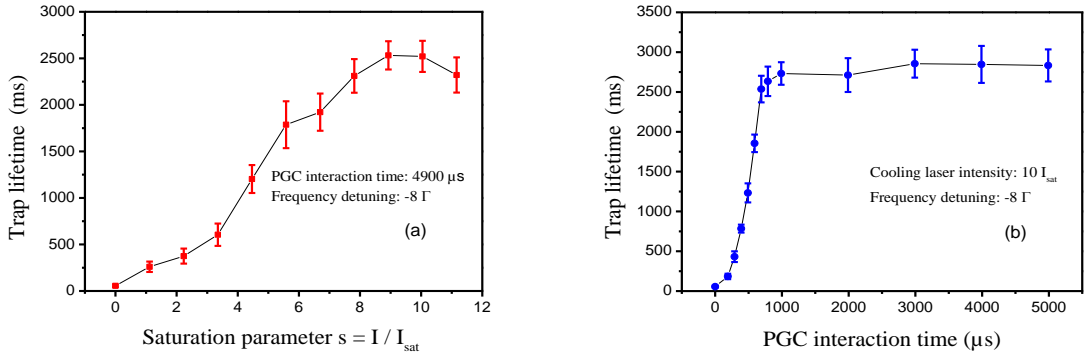

**FIG. 4** Trap lifetime as a function of the cooling laser intensity (a) and the PGC interaction time (b). (a) shows that when the cooling laser intensity is weak, the cooling effect is insufficient. The cooling laser intensity is changed from 0 to 12Isat. (b) shows that the trap lifetime increases rapidly from ~ 56 ms to ~ 2536 ms as the cooing time increases from ~ 190 μs to ~ 690 μs, and reaches an almost steady value after 690 μs. Experimental parameters: pulse duration: 5 ns, repetition rate: 10 MHz, pulse laser beam's power: ~ 1.25 mW, pulse laser beam's waist: ~ 12 μm.

To demonstrate the viability of the suppression of single-atom heating for pulsed excitation experiments, we measure the trap lifetime with an optimized gated pulse excitation and cooling technique. The single atom is excited in the radial direction and the repetition rate of the pulsed laser is 10 MHz. The single atom is excited for 10 μs and cooled for 690 μs and the cooling laser intensity is 10 $I_{sat}$. By measuring the recapture probability as a function of the pulsed excitation time, we obtain a trap lifetime of τ = 2536 ± 31 ms (Fig. 5). The trap lifetime in the absence of cooling is limited by the repetition rate of the pulsed laser and the trap frequency. The heating due to the momentum accumulation process increase the energy of the atom. The trap lifetime also can be improved by increase the trap depth. The result shows that the heating rate is greatly suppressed. The long trap lifetime shown here also allow us to obtain a long time for pulsed excitation. The corresponding number of allowed excitation is increased from ~ 108 to ~ 360000. The optimized excitation and cooling sequence enables us to suppress the heating and obtain the maximum excitation time of the single atom. It is particularly important for the application of triggered single-photon source, since the maximum number of single photons is desired.

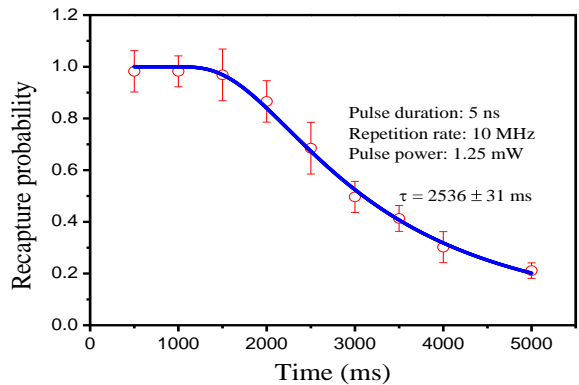

**FIG. 5** Measured recapture probability (red circles) as a function of the duration time of the excitation pulse laser. The blue line is fitting to the data. The cooling laser intensity is 8Isat and the PGC interaction time is 690 μs. By optimizing the parameters of the cooling laser, the trap lifetime is extended to 2536 ± 31 ms. The numbers of excitation have been improved from ~ 108 to ~ 360000. The trap lifetime is limited by the atomic heating due to the momentum accumulation process induced by the pulse excitation. Experimental parameters: pulse duration: 5 ns, repetition rate: 10 MHz, pulse laser beam's power: 1.25 mW, pulse laser beam's waist: ~ 12 μm, $I_{cool}$ = 10 $I_{sat}$, PGC time: 690 μs.

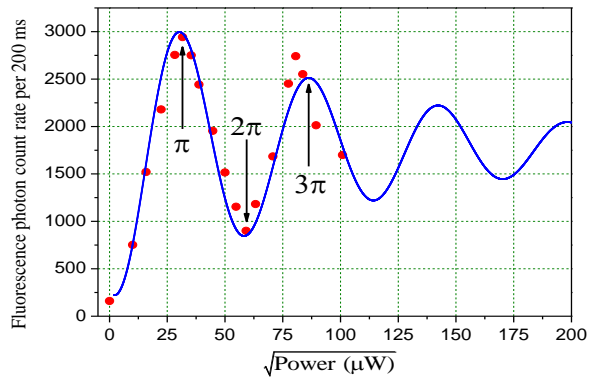

**FIG. 6** Fluorescence photon count rate (red dots) as a function of the square root of the pulse laser power. The laser pulse duration is fixed at 5 ns, with a repetition rate of 10 MHz. The blue solid line is a theoretical curve based on a simple two-level model. The π pulsed laser power is ~ 1.25 mW with pulse laser beam's waist: ~ 12 μm.

In the following, we describe the use of the cooled single atom to generate triggered single photons. The pulsed laser is used to drive Rabi oscillations on the Cs $6S_{1/2}$ |Fg = 4, $m_F$ = +4> (ground state) - $6P_{3/2}$ |Fe = 5, $m_F$ = +5> (excited state) cycling transition. The atom can continue to oscillate between the ground state and the excited state as long as it is interacting with the excitation light. A square resonant pulse with correct duration and power can transfer the atom from the ground state to the excited state. This is referred to as the π pulse. As shown in Fig. 6, the fluorescence photon count rate detected by the SPCM is plotted after the pulse excitation as a function of the square root of the pulse laser power. The laser pulse duration is fixed at 5 ns, with a repetition rate of 10 MHz. The Rabi oscillations are visible in the results. The π pulsed laser power is ~ 1.25 mW. During the pulsed excitation process, the single-photon collection rate into the fiber is ~ 29820 photons/s, which corresponds to a collection efficiency of ~ 0.3%. In addition, we see that for 2π-pulses, the excitation probability does not decrease to zero, but stays at a finite value. This reflects the finite probability of emitting a photon during the excitation pulse. The fluctuations of the pulsed laser peak power lead to a reduction in the contrast of the oscillations at a higher laser power [10].

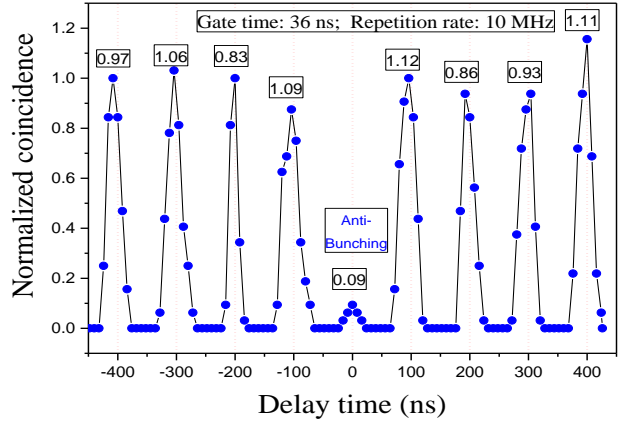

**FIG. 7** Second-order intensity correlations of the single-photon source with background subtraction. Black dots are experimental data with 8 ns time bin. The absence of a peak at zero delay shows that the source is emitting single photons. Experimental parameters: pulse duration: 5 ns, repetition rate: 10 MHz, pulse laser beam's power: ~ 1.25 mW, pulse laser beam's waist: ~ 12 μm, $I_{cool} = 10I_{sat}$, PGC time: 690 μs.

After π-pulse excitation, a photon will be generated owing to the spontaneous atomic decay to the ground state. By repeating these steps with high repetition rate, we expected to generate a triggered single photon on demand. To prove the single-photon characteristics of the source, the photon correlation function is measured by implementing a HBT setup [18], which contains information on photon emission statistics. For a pulsed excitation, the second-order correlation function $g^{(2)}(\tau)$ becomes a series of peaks separated by the laser repetition period, and the areas of these peaks give information on photon number correlations between pulses separated by time τ. As the experimental setup shows in Fig. 1, the P7888 card records the stop events during 1 μs after a start pulse with a time resolution of 8 ns. To investigate negative correlation times, a constant delay (500 ns) is introduced in the stop channel. The results are presented in Fig. 7, $g^{(2)}(\tau)$ is measured from correlations between photon arrival times at the two SPCMs as a function of time delay τ. The background is subtracted which mainly from the SPCMs dark counts, when no atom is trapped, and the rest come from various sources of scattered light. In order to reduce the background level, the detection gate is used so that only photons scattered during the periods of pulsed excitation are counted. The result is normalized in the following way. The normalized area of each peak is given by: $C_N(m) = c(m)/(N_1 N_2 \theta T)$ [26], where $c(m)$ is the area of the peak $m$, $N_{1;2}$ are the count rates on each SPCM, $\theta$ is the repetition period and $T$ is the total pulsed excitation time. The peaks at multiples of 100 ns delay are owing to correlations between photons generated by different excitation pulses. The defining characteristic of a single photon source is evident in the absence of a peak at τ= 0，which reflects the upper bound of the probability of detecting two photons. Dividing the total integrated residual area within a 72 ns window around zero delay by the average area of the adjacent eight peaks, gives approximately the probability to emit two photons per excitation pulse [27]. The multiphoton probability is calculated to be $g^2(0)$ = 0.09, whereas the probability due to the correlation between the single photon and dark count noise is about ~5%. The probability of double excitation is ~4% owing to a small probability for the atom to emit a photon during the pulse, and be re-excited and emit a second photon [10, 11].

## V. CONCLUSION

In conclusion, we have investigated the heating of a trapped single atom induced by the pulsed excitation in a microscopic optical dipole trap. The heating depends on the trap frequency and it limits the maximum repetition rate of the pulsed laser. We have found that, after applying a gated excitation and cooling method, the heating of the trapped atom is effectively suppressed. We experimentally demonstrate this method with optimized cooling parameters. The trapped atom can be excited over 2536 ms. The corresponding number of excitations have been improved from ~ 108 to ~ 360000. Finally, we use the cooled single atom as a single-photon source and have measured the second-order correlations of the emitted 852nm single photons by implementing a HBT setup. The rate of 10 MHz of single-photon source is significantly higher than the repetition rate restriction by the trap oscillation frequency. The next, we would like to use the 935.6 nm magic wavelength optical dipole trap for cesium [28] to suppress the broadening induced by the differential light shifts. Moreover, this method can also be used to suppress the phase decoherence in an atomic qubit and demonstrate high-speed quantum logic, ultrafast single qubit or multiqubit operations [6-9].

**ACKNOWLEDGMENTS**

This research work is financially supported by the National Natural Science Foundation of China (Grant Nos. 11274213, 61475091, and 61205215) and the National Major Scientific Research Program of China (Grant No. 2012CB921601).